\documentclass[12pt,oneside]{article}

\usepackage[margin=1in]{geometry}
\usepackage{setspace}
\usepackage{amsmath}
\usepackage{amssymb}
\usepackage{amsthm}
\usepackage{exscale}
\usepackage[mathscr]{eucal}
\usepackage{bm}
\usepackage{bbm}
\usepackage{eqlist} 
\usepackage[final]{graphicx}
\usepackage[dvipsnames]{color}
\usepackage{verbatim}
\usepackage{natbib}
\usepackage[justification=centering]{caption}
\usepackage{enumerate}
\usepackage{algorithm}
\usepackage{algorithmic}
\usepackage{dsfont}
\usepackage{diagbox}
\usepackage{rotating}
\usepackage{lscape}
\usepackage{multirow}
\usepackage{rotating}
\usepackage[table,xcdraw]{xcolor}
\usepackage{subfigure}
\usepackage{url}
\usepackage{hyperref}
\hypersetup{
    colorlinks=FALSE,
    linkcolor=blue,
    filecolor=magenta,
    urlcolor=cyan,
}

\usepackage{etoolbox}
\usepackage{booktabs}

\makeatletter
\patchcmd{\env@cases}{1.2}{0.6}{}{}
\makeatother

\DeclareGraphicsExtensions{.pdf, .jpg}
\DeclareMathOperator*{\argmax}{arg\,max}

\theoremstyle{definition}

\title{\large Model-Based Clustering of Nonparametric Weighted Networks with Application to Water Pollution Analysis}
\author{\normalsize Amal Agarwal and Lingzhou Xue \\ \normalsize Department of Statistics, Pennsylvania State University \\ \normalsize}
\date{}
\begin{document}
\maketitle
\abstract{Water pollution is a major global environmental problem, and it poses a great environmental risk to public health and biological diversity. This work is motivated by assessing the potential environmental threat of coal mining through increased sulfate concentrations in river networks, which do not belong to any simple parametric distribution. However, existing network models mainly focus on binary or discrete networks and weighted networks with known parametric weight distributions. We propose a principled nonparametric weighted network model based on exponential-family random graph models and local likelihood estimation, and study its model-based clustering with application to large-scale water pollution network analysis. We do not require any parametric distribution assumption on network weights. The proposed method greatly extends the methodology and applicability of statistical network models. Furthermore, it is scalable to large and complex networks in large-scale environmental studies. The power of our proposed methods is demonstrated in simulation studies and a real application to sulfate pollution network analysis in Ohio watershed located in Pennsylvania, United States.
}

\medskip
{\bf Keywords:} Exponential-Family Random Graphical Model, Local Likelihood, Variational Inference, Environmental Studies.

\newpage

\section{Introduction}

Water pollution is the leading cause of deaths and diseases, and it is a major global problem. It is known that nearly $80\%$ of the world's population lives in areas exposed to high levels of threat to water security \citep{vorosmarty2010global}.  The recent national report on water quality by \cite{epa2017} pointed out that $46\%$ of rivers, $21\%$ of lakes, $18\%$ of coastal waters, and $32\%$ of wetlands in the United States are in poor biological condition or rated poor based on a water quality index. The major pollutant sources include agriculture, atmospheric deposition, construction, industrial production, municipal sewage, resource extraction, spills, and urban runoff. They pose severe health hazards like cancer, cardiovascular, respiratory, neurologic, and developmental damage. This work is motivated by assessing the potential environmental threat of coal mining in Ohio watershed of Pennsylvania through increased sulfate concentrations in the surface water, which is an important scientific problem in geoscience. \cite{bernhardt2012many} mapped surface coal mining of southern West Virginia and linked these maps with water quality and biological data of 223 streams. When these coal mines occupy $>5.4\%$ of their contributing watershed area, the sulfate concentrations within catchments could exceed $50$ mg/L \citep{niu2017detecting}. Residential proximity to heavy coal production is associated with higher risk for cardiopulmonary disease, chronic lung disease, hypertension, and kidney disease \citep{hendryx2008relations}. The study of water-quality risks will help the whole society manage them now and in future.

With advances in data collection, there are more and more modern statistical research on environmental studies using nonparametric regression, causal inference, mixture model, network analysis, and variable selection, for instance, \cite{ebenstein2012consequences,liang2015assessing,li2017discovery,lin2017statistical,wen2019assessing} among many others.
Especially, network analysis becomes increasingly important in large-scale environmental studies and geoscientific research to assess environmental impacts and
risks for water pollution \citep{smith1987water,lienert2013stakeholder,ruzol2017understanding}. For example,, \cite{gianessi1981analysis} proposed a water network model to explore the impact of cropland sediment controls on improved water quality, and \cite{montgomery1972markets} and \cite{anastasiadis2016network} studies weighted pollution networks where the weights measure the pollution diminishing transition. However, none of aforementioned network models took into account the spatial heterogeneity and the hub structure of river networks.
Without exploring spatial heterogeneity, these models could fail to differentiate polluted regions from less polluted but well connected regions in river networks.
The hubs in river networks usually determine the flow of pollutants, and they may help identify polluted and well connected regions.

In this work, we introduce a principled model-based clustering of networks to effectively deal with the spatial heterogeneity of river pollution and efficiently identify the hub structure in river networks. Model-based clustering of networks based on stochastic block models (SBMs) and exponential-family random graph models (ERGMs) have received considerable attention in recent literature, including \cite{snijders-1997,nowicki-2001,girvan-newman-2002,airoldi2008mixed,karrer-newman-2011,zhao2012consistency,vu2013model,saldana2017many,wang2017likelihood,lee2017model}, among many others. It is worth pointing out that existing research mainly focuses on the model-based clustering of networks with binary or discrete edges. In a recent paper by \cite{ambroise2012new}, parametric distributions are incorporated into a stochastic block model to model continuous network weights. Alternatively, Bayesian variational methods have been proposed by \cite{aicher2014learning} to approximate posterior distribution of weights over latent block structures. However, from our motivating example, sulfate concentrations in the surface river network do not belong to any simple parametric distribution, which will be illustrated in Section 5. Thus, we need to relax the parametric assumption in network models to account for the unknown distribution of continuous network weights. To address this issue, we propose a new nonparametric weighted network model based on ERGMs and local likelihood estimation, and study its model-based clustering with application to large-scale water pollution network analysis. The proposed method greatly extends the methodology and applicability of statistical network models. Furthermore, it is scalable to large and complex networks in real-world applications.

The rest of this paper is organized as follows. Section 2 presents the methodology of our proposed nonparametric weighted network model. In Section 3, we introduce a novel variational expectation-maximization algorithm to solve the approximate maximum likelihood estimation.  Section 4 demonstrates the numerical performances of our proposed methods and algorithms in simulation studies. In Section 5, we apply the proposed method to analyze the large-scale water pollution network of sulfate concentrations in the Ohio watershed of Pennsylvania.

\section{Methodology}
We define some necessary notation before presenting our proposed method. Let $n$ be the number of nodes in the observed network.  Let $\mathbf{Y}=(Y_{ij})_{1\leq i,j\leq n}$ be the corresponding weighted network such that $Y_{ij}=(E_{ij},W_{ij})$ where $E_{ij}$ is a binary indicator denoting the existence of an edge in the network and $W_{ij}$ is the corresponding weight when $E_{ij}=1$. The weight matrix $W=(W_{ij})_{1\leq i,j\leq n}$ consists of continuous weights in the network. Further we assume that the edge distribution of each network belongs to an exponential family \citep{besag1974spatial,frank1986markov}. We write the distribution of edge indicator matrix $\boldsymbol{E}$ as
\begin{equation}\label{netexp}
	P_{\boldsymbol{\theta}}(\boldsymbol{E} = \boldsymbol{e}) = \mbox{exp}\{\boldsymbol{\theta}'\boldsymbol{g}(\boldsymbol{e})-\psi(\boldsymbol{\theta})\},
\end{equation}
where $\psi(\boldsymbol{\theta}) = \mbox{log}\sum_{\boldsymbol{e} \in \mathcal{E}}\mbox{exp}\left[\boldsymbol{\theta}'\boldsymbol{g}(\boldsymbol{e}) \right]$ is the log of the normalizing constant, $\boldsymbol{\theta}\in \mathbb{R}^{p}$ are canonical network parameters of interest and $\boldsymbol{g}: \mathcal{E}\rightarrow \mathbb{R}^{p}$ is the sufficient statistic. Here $\mathcal{E}$ is the space for $\boldsymbol{E}$ consisting of $2^n$  possible binary edge structures.

One of the major limitations of this binary network model is that it can not deal with large number of nodes due to large computational time for evaluating the likelihood function. For undirected networks, this computing time scales with node size as $\exp{((n(n-1)\log{2})/2)}$. Many estimation algorithms have been developed \citep{snijders2002markov,hunter2006inference,moller2006efficient,koskinen2010analysing,caimo2011bayesian}, however most of them are time-consuming and therefore unrealistic for fitting large networks. This issue of non-scalability can be resolved by the assumption of dyadic independence which assumes that all dyads are independent of each other. Note that dyad is a general term applicable for both directed and undirected edges. In the undirected weighted networks, it would imply the ties and weights are independent of each other and for all pairs of nodes, i.e.
\begin{equation}\label{dyadIndep}
	P_{\boldsymbol{\theta}}(\boldsymbol{Y} = \boldsymbol{y}) = \prod_{1\leq i<j\leq n}\mbox{exp}\{\boldsymbol{\theta}'\boldsymbol{g}(e_{ij})-\psi(\boldsymbol{\theta})\}P_{\boldsymbol{\theta}}({W}_{ij} = {w}_{ij}),
\end{equation}
This assumption facilitates both estimation and simulation of large networks as well as solves the issue of degeneracy \citep{strauss1986general,handcock2003assessing,schweinberger2011instability,krivitsky2012exponential}.
However, dyadic independence is too restrictive and most models following this assumption are either very trivial, failing to capture relational dependencies \citep{gilbert1959random,erdds1959random} or non-parsimonious, with a large number of parameters \citep{holland1981exponential}.

We consider a model-based clustering framework to relax the dyadic independence. More specifically, we introduce the finite $K$-component mixture form together with a much less restrictive assumption of \emph{conditional dyadic independence} (CDI) \citep{snijders-1997,nowicki-2001,girvan-newman-2002,vu2013model}. Under this assumption, we propose the nonparametric weighted network as
\begin{equation}\label{cdi}
    	P_{\boldsymbol{\theta},\boldsymbol{f}}(\mathbf{Y} = \mathbf{y} | \mathbf{Z}=\mathbf{z}) =\prod_{1\le i<j\le n}P_{\boldsymbol{\theta}_{z_{i}z_{j}},{f}_{z_{i}z_{j}}}(Y_{ij}=y_{ij}|\boldsymbol{Z}=\boldsymbol{z})
\end{equation}
where $\mathbf{Z} = (\boldsymbol{Z}_{i})_{1\leq i\leq n}$ denote the latent cluster memberships of nodes.
Here, $\boldsymbol{Z}_{i}$ is a $K\times 1$ vector such that $z_{ik}=1$ if and only if node $i$ lies in cluster $k$, otherwise $z_{ik}=0$. The conditional dyadic independence strikes in a nice balance between model complexity and parsimony for model-based clustering. The number of parameters are reduced from $\mathcal{O}(n^2)$ to $\mathcal{O}(K^2)$, thus enabling simple inter and intra community interpretations. The CDI assumption induces a block structure similar to SBMs.  SBMs have been thoroughly studied in the context of social networks \citep{holland1983stochastic,airoldi2008mixed} and have a long history in multiple scientific communities \citep{bui1987graph,dyer1989solution,bollobas2007phase}. We omit discussion of SBMs here except to point out a major difference from our current setup of ERGM. ERGMs can allow several kinds of dynamic network statistics like density, stability, transitivity \citep{hanneke2010discrete}, thus effectively generalizing the simple density case in SBM methodology which makes them appealing in practice. To effectively model continuous weights, for any given pair of nodes $(i,j)$, we have
\begin{equation}\label{psnd}
	\begin{split}
		P_{\boldsymbol{\theta},\boldsymbol{f}}(Y_{ij}=y_{ij}|\boldsymbol{Z}=\boldsymbol{z}) & = \left(p_{z_iz_j}f_{z_iz_j}(w_{ij})\right)^{\mathbbm{1}_{e_{ij}\neq 0}}\left(1-p_{z_iz_j}\right)^{\mathbbm{1}_{e_{ij}=0}}
	\end{split}
\end{equation}
Here, $(p_{kl})_{1\leq k,l\leq K}=(P_{\boldsymbol{\theta}_{z_{i}z_{j}}}(E_{ij} = e_{ij} | \mathbf{Z}=\mathbf{z}))_{1\leq k,l\leq K}$ take the parametric specification of exponential-family distributions of the network statistics as explained in \eqref{netexp} and network weights $(w_{ij})_{1\leq i,j\leq n: z_{i}=k, z_{j}=l}$ are assumed to be an \emph{i.i.d.} sample observed from a population following an univariate nonparametric density function $f_{kl}$. Note that $(f_{kl})_{1\leq k,l\leq K}$ do not necessarily have any parametric form. We implicitly assume that conditioning on the full $Z = z$ is same as conditioning on just $z_i$ and $z_j$, i.e. $E_{ij}$ and $W_{ij}$ depend on $Z$ only via $z_i$ and $z_j$. For ease of presentation, we assume the additive structure of $p_{kl}$ to be parameterized by network sparsity parameters $\boldsymbol{\theta}$ as $p_{kl}=\text{logit}^{-1}{(\theta_k+\theta_l)}$. It is worth pointing out that  the clusters $z_1,\cdots,z_n$ are determined by two sources of information in network models. On the one hand, the random block structure and also the additive structure of $(p_{kl})_{1\leq k,l\leq K}$ contribute to the exploration of different degrees among the clusters. On the other hand, the nonparametric network weights modulate the separation between clusters with different weight distributions.


Combining \eqref{cdi} and \eqref{psnd}, the corresponding log-likelihood function given the cluster memberships can be written as:
\begin{equation}\label{cond_log_lik}
    \begin{split}
        &\log(P_{\boldsymbol{\theta},\boldsymbol{f}}(\mathbf{Y}=\mathbf{y}\mid\mathbf{Z}=\mathbf{z}))\\=&\sum_{i=1}^{n-1}\sum_{j=i+1}^n\Bigg\{\big[\mathbbm{1}_{e_{ij}\neq 0}\log(p_{z_i z_j})+\mathbbm{1}_{e_{ij}= 0}\log(1-p_{z_i z_j})\big]\\
        &+\mathbbm{1}_{e_{ij}\neq 0}\Bigg[\log(f_{z_i z_j}(w_{ij}))-\Big(\int_{\mathcal{X}}f_{z_i z_j}(u)du-1\Big)\Bigg]\Bigg\},
    \end{split}
\end{equation}
where we also introduce the penalty term $\Big(\int_{\mathcal{X}}f_{kl}(u)du-1\Big)$. Thus, \eqref{cond_log_lik} can be treated as a likelihood for any non-negative function $f_{kl}$ without imposing the additional constraint $\int_{\mathcal{X}}f_{kl}(u)du=1$. This specification follows the similar spirit of \cite{loader1996local}.

Now we derive the localized version of the conditional log-likelihood evaluated at an arbirary grid point $w$ as:
\begin{equation}\label{local_cond_log_lik}
    \begin{split}
        &\ell(\boldsymbol{\theta},\boldsymbol{f},w;\boldsymbol{Y}\mid\boldsymbol{Z})\\
        =&\sum_{i=1}^{n-1}\sum_{j=i+1}^n\Bigg\{\left[\mathbbm{1}_{e_{ij}\neq 0}\log(p_{z_i z_j})+\mathbbm{1}_{e_{ij}= 0}\log(1-p_{z_i z_j})\right]+\mathbbm{1}_{e_{ij}\neq 0}\times\\&\Bigg[K_h(w_{ij}-w)\log(f_{z_i z_j}(w_{ij}))-\Big(\int_{\mathcal{X}}K_h(u-w)f_{z_i z_j}(u)du-1\Big)\Bigg]\Bigg\}
    \end{split}
\end{equation}
where $K_h$ is the rescaled kernel function with a positive bandwidth $h$. We approximate $\log(f_{kl}(u))$ by $\Phi^p_{kl}$, a linear combination of orthogonal basis functions $(\phi_m)_{1\leq m\leq p}$, namely, $\Phi^p_{kl}(u-w)=\sum_{m=0}^p\beta_m^{(kl)}\phi_m(u-w)$. With this approximation, the local conditional log-likelihood becomes
\begin{equation}\label{local_phi_approx_cond_log_lik}
    \begin{split}
        &\ell(\boldsymbol{\theta},\boldsymbol{\beta},w;\boldsymbol{Y}\mid\boldsymbol{Z})\\=&\sum_{i=1}^{n-1}\sum_{j=i+1}^n\Bigg\{\big[\mathbbm{1}_{e_{ij}\neq 0}\log(p_{z_i z_j})+\mathbbm{1}_{e_{ij}= 0}\log(1-p_{z_i z_j})\big]+\mathbbm{1}_{e_{ij}\neq 0}\times\\
        &\Bigg[K_h(w_{ij}-w)\Phi^p_{z_i z_j}(w_{ij}-w)-\Big(\int_{\mathcal{X}}K_h(u-w)\exp(\Phi^p_{z_i z_j}(u-w))du-1\Big)\Bigg]\Bigg\}
    \end{split}
\end{equation}

We assume that membership indicators $\mathbf{Z} = (Z_i)_{1\leq i\leq n}$ follow a multinomial distribution with a single trial and mixture proportions as $\mathbf{\pi}=(\pi_k)_{1\leq k\leq K}$. The log-likelihood of the observed weighted network can be written as
\begin{equation}\label{marg_log_lik}
    \ell(\boldsymbol{\theta},\boldsymbol{\pi},\boldsymbol{f})=\log\Bigg(\sum_{\mathbf{z}\in\{1,\ldots,K\}^n}P_{\boldsymbol{\theta},\boldsymbol{f}}(\mathbf{Y}=\mathbf{y}\mid\mathbf{Z}=\mathbf{z})P_{\boldsymbol{\pi}}(\mathbf{Z}=\mathbf{z})\Bigg)
\end{equation}

In view of \eqref{cond_log_lik} and \eqref{local_phi_approx_cond_log_lik}, we maximize the log-likelihood function \eqref{marg_log_lik} of the observed network described to estimate model parameters $\boldsymbol{\theta}$ together with block densities $\boldsymbol{f}$.

\textbf{Remark 1.} \cite{allman2011parameter} proved the identifiability of parameters for a broad class of parametric or nonparametric weighted network models. As shown in Section 4 of \cite{allman2011parameter}, we can uniquely identify the parameters under mild conditions for parametric weighted networks while the identifiability result for nonparametric weighted networks depend on binning the values of the edge variables into a finite set. Please see Theorem 15 of \cite{allman2011parameter} for more details about the identifiability.

\textbf{Remark 2.} The proposed nonparametric  weighted network model can be further extended to discrete temporally evolving weighted networks. Given the dynamic network series $\mathbf{Y}=(\mathbf{Y}_t)_{1\leq t\leq T}=(\mathbf{E}_t,\mathbf{W}_t)_{1\leq t\leq T}$, the dynamic nonparametric weighted network model can be derived by assuming a discrete-time Markov structure over the time \citep{hanneke2010discrete,krivitsky2014separable,kim2018review}, that is
$$P_{\boldsymbol{\theta},\boldsymbol{f}}(\mathbf{Y} = \mathbf{y} | \mathbf{Z}=\mathbf{z}) =\prod_{2\le t\le T}P_{\boldsymbol{\theta},\boldsymbol{f}}(\mathbf{Y}_t = \mathbf{y}_t | \mathbf{Y}_{t-1} = \mathbf{y}_{t-1},  \mathbf{Z}=\mathbf{z})$$
where $P_{\boldsymbol{\theta},\boldsymbol{f}}(\mathbf{Y}_t = \mathbf{y}_t | \mathbf{Y}_{t-1} = \mathbf{y}_{t-1},  \mathbf{Z}=\mathbf{z})$ follows the similar specification of \eqref{cdi} and \eqref{psnd}. We can incorporate dynamic network statistics such as stability and transitivity in the exponential-family specification of $P_{\boldsymbol{\theta}_{z_{i}z_{j}}}(E_{t,ij} = e_{t,ij}\mid E_{t-1,ij} = e_{t-1,ij},\mathbf{Z}=\mathbf{z})$. Hence, the interpretation of clusters will reflect the impact from dynamic network statistics. For example, the use of stability network statistic will contribute to the exploration of different levels of stability among the clusters. The dynamic nonparametric weighted network is beyond our current scope and we will study it in the future.

\section{Computation}

This section proposes a variational expectation maximization (EM) algorithm to approximately solve the maximum likelihood estimation. It is infeasible to directly maximize the log-likelihood function \eqref{marg_log_lik} due to two key challenges: (i) exponential-family form of $P(\boldsymbol{Y} \mid \boldsymbol{Z})$ is not scalable for large networks; (ii) the sum is over every possible assignment to $\boldsymbol{Z}$, where for each $1\le i\le n$ , $Z_{i}=z_{i}$ can take one of $K$ possible values. 

To resolve the first challenge, the CDI assumption \eqref{cdi} plays a crucial role. Typically parameters in a mixture model are estimated using the classical EM algorithm \citep{dempster1977maximum}. The E-step proceeds by writing the complete data log-likelihood \eqref{complete_log_lik}, assuming the network is observed while node membership indicators $\boldsymbol{Z}$ are unobserved.
\begin{equation}\label{complete_log_lik}
	\begin{split}
		&\log(P_{\boldsymbol{\theta},\boldsymbol{\pi},\boldsymbol{f}}(\mathbf{Y}=\mathbf{y},\mathbf{Z}=\mathbf{z})) \\=& \sum_{i=1}^{n-1}\sum_{j =i+1}^n\sum_{k=1}^{K}\sum_{l=1}^{K}z_{ik}z_{jl}\log P_{\mathbf{\theta}}(Y_{ij}=y_{ij} \mid \mathbf{Z}=\mathbf{z})+\sum_{i=1}^{n}\sum_{k=1}^{K}z_{ik}\log(\pi_k)
	\end{split}
\end{equation}
Next we take expectation of this complete log-likelihood with the distribution $\mathbb{P}_{\boldsymbol{\theta}}(\boldsymbol{Z}\mid\boldsymbol{Y})$. Clearly the distribution $\mathbb{P}_{\boldsymbol{\theta}}(\boldsymbol{Z}\mid\boldsymbol{Y})=\mathbb{P}_{\boldsymbol{\theta}}(\boldsymbol{Z}_{i, 1\le i\le n}\mid\boldsymbol{Y})$ can't be further factored over nodes since for $i\ne j$, $Z_{i}$ is not independent of $Z_{j}$ given the observed network $\boldsymbol{Y}$. This poses a huge computational challenge.
The intractability of the complete log-likelihood motivates the use of variational approximation. Variational methods are well studied in literature \citep{blei2017variational}. The basic idea is to posit a tractable auxiliary distribution $A_{\gamma}(\boldsymbol{z}) \equiv P(\boldsymbol{Z} = \boldsymbol{z})$ for the latent variables $\boldsymbol{Z}$ and find the optimal setting for variational parameters $\gamma$ that minimizes the Kullback-Liebler divergence between the approximation and true distribution. We use this auxiliary distribution to construct a tractable lower bound of the log-likelihood using Jensen's inequality and then maximize this lower bound, yielding approximate maximum likelihood estimates.
\begin{equation}\label{tracl}
	\begin{split}
		\ell(\boldsymbol{\theta},\boldsymbol{\pi},\boldsymbol{f})
		= & \log \left(\sum_{\mathbf{z} \in \{1,\ldots,K\}^n}\dfrac{P_{\boldsymbol{\theta},\boldsymbol{f}}(\mathbf{Y} = \mathbf{y} | \mathbf{z})P_{\boldsymbol{\pi}}(\mathbf{Z} = \mathbf{z})}{A_{\boldsymbol{\gamma}}(\mathbf{z})}A_{\boldsymbol{\gamma}}(\mathbf{z})\right)\\
		\ge & \sum_{\mathbf{z} \in \{1,\ldots,K\}^n}\log\left(\dfrac{P_{\boldsymbol{\theta},\boldsymbol{f}}(\mathbf{Y} = \mathbf{y}\mid \mathbf{z})P_{\boldsymbol{\pi}}(\mathbf{Z} = \mathbf{z})}{A_{\boldsymbol{\gamma}}(\mathbf{z})}\right)A_{\boldsymbol{\gamma}}(\mathbf{z})\\
		= & \text{ ELBO }(\boldsymbol{\theta},\boldsymbol{\gamma},\boldsymbol{\pi},\boldsymbol{f})
	\end{split}
\end{equation}
The derivation in \eqref{tracl} uses an auxiliary distribution and also Jensen's inequality. We choose the variational distribution $A(\mathbf{Z})$ from the mean-field family as, 
\begin{equation}\label{az}
	\begin{split}
		A_{\boldsymbol{\gamma}}(\mathbf{Z}) & = \prod_{i=1}^{n}P_{\boldsymbol{\gamma}_{i}}(\mathbf{Z}_{i})  = \prod_{i=1}^{n}\prod_{k=1}^{K}\gamma_{ik}^{z_{ik}}
	\end{split}
\end{equation}
where $\forall$ $i \in \{1,\dots,n\}$, $k \in \{1,\dots,K\}$, we have $0 \le \gamma_{ik} \le 1$ with the constraint $\sum_{k=1}^{K}\gamma_{ik}=1$ and $z_{ik}=\mathbbm{1}_{z_i=k}$. This class of probability distributions $A_{\boldsymbol{\gamma}}$ considers independent laws through different nodal memberships. With the definition of $A(\mathbf{Z})$ in \eqref{az}, we derive the following effective lower bound (ELBO).
\begin{equation}\label{ELBO}
    \begin{split}
        &\text{ ELBO }(\boldsymbol{\theta},\boldsymbol{\gamma},\boldsymbol{\pi},\boldsymbol{f})\\= & \sum_{i = 1}^N\sum_{j = i+1}^N\sum_{k=1}^{K}\sum_{l=1}^{K}\Bigg\{\gamma_{ik}\gamma_{jl}\big[\mathbbm{1}_{e_{ij}\neq 0}\log(p_{kl})+\mathbbm{1}_{e_{ij}= 0}\log(1-p_{kl})\big]\\
        &+\mathbbm{1}_{e_{ij}\neq 0}\Bigg[\log(f_{kl}(w_{ij}))-\Big(\int_{\mathcal{X}}f_{kl}(u)du-1\Big)\Bigg]\Bigg\} \\
		&+ \sum_{i=1}^{N}\sum_{k=1}^{K}\gamma_{ik}\left[\log(\pi_{k})-\log(\gamma_{ik})\right]
    \end{split}
\end{equation}

\textbf{Variational E-step}: we maximize $\text{ELBO }(\boldsymbol{\theta},\boldsymbol{\gamma},\boldsymbol{\pi},\boldsymbol{f})$ in \eqref{ELBO} to obtain $\boldsymbol{\gamma}^{(t)}$:
\begin{equation}\label{gamma_update}
	\boldsymbol{\gamma}^{(t)}=\argmax_{\boldsymbol{\gamma}}\text{ ELBO }(\boldsymbol{\theta}^{(t-1)},\boldsymbol{\gamma},\boldsymbol{\pi}^{(t-1)},\boldsymbol{f}^{(t-1)})
\end{equation}
The direct maximization of ELBO in \eqref{gamma_update} is difficult, since the lower bound depends on the products $\gamma_{ik}\gamma_{jl}$ and, therefore the fixed-point updates of $\gamma_{ik}$ depend on all other $\gamma_{jl}$ \citep{daudin2008mixture}. To separate the parameters in this maximization problem, we adopt an MM algorithm that involves constructing a surrogate (minorizing) function and optimizing it iteratively \citep{hunter2004tutorial}. The surrogate function Q must satisfy the following properties to qualify as a valid minorizing function.
\begin{equation}\label{surrogate_property_1}
	Q(\boldsymbol{\theta}^{(t)},\boldsymbol{\gamma}^{(t)},\boldsymbol{\pi}^{(t)},\boldsymbol{f}^{(t)};\boldsymbol{\gamma})\leq \text{ELBO}(\boldsymbol{\theta}^{(t)},\boldsymbol{\gamma},\boldsymbol{\pi}^{(t)},\boldsymbol{f}^{(t)})\hspace{0.1in} ,\forall \boldsymbol{\gamma}
\end{equation}
\vspace{-0.4in}
\begin{equation}\label{surrogate_property_2}
	Q(\boldsymbol{\theta}^{(t)},\boldsymbol{\gamma}^{(t)},\boldsymbol{\pi}^{(t)},\boldsymbol{f}^{(t)};\boldsymbol{\gamma}^{(t)})= \text{ELBO}(\boldsymbol{\theta}^{(t)},\boldsymbol{\gamma}^{(t)},\boldsymbol{\pi}^{(t)},\boldsymbol{f}^{(t)})
\end{equation}

First we note that for all $(\theta_{kl})_{1\leq k\leq l\leq K}$ we have $\log(p_{kl}) < 0$ and $\log(1-p_{kl}) < 0$ which gives rise to following inequalties using the arithmetic geometric mean inequality:
\begin{equation}\label{p_kl_inequality} \gamma_{ik}\gamma_{jl}\log(p_{kl})\geq\Big(\gamma_{ik}^2\dfrac{\hat{\gamma}_{jl}}{2\hat{\gamma}_{ik}}+\gamma_{jl}^2\dfrac{\hat{\gamma}_{ik}}{2\hat{\gamma}_{jl}}\Big)\log(p_{kl})
\end{equation}
\vspace{-0.4in}
\begin{equation}\label{1-p_kl_inequality} \gamma_{ik}\gamma_{jl}\log(1-p_{kl})\geq\Big(\gamma_{ik}^2\dfrac{\hat{\gamma}_{jl}}{2\hat{\gamma}_{ik}}+\gamma_{jl}^2\dfrac{\hat{\gamma}_{ik}}{2\hat{\gamma}_{jl}}\Big)\log(1-p_{kl})
\end{equation}
with equality if $\gamma_{ik}=\hat{\gamma}_{ik}$ and $\gamma_{jl}=\hat{\gamma}_{jl}$. Also the concavity of the logarithm function gives rise to following inequality \citep{vu2013model}
\begin{equation}\label{log_inequality}
	-\log(\gamma_{ik})\geq-\log(\hat{\gamma}_{ik})-\dfrac{\gamma_{ik}}{\hat{\gamma}_{ik}}+1
\end{equation}

We construct the surrogate function that satisfies \eqref{surrogate_property_1} and \eqref{surrogate_property_2} using the inequalities \eqref{p_kl_inequality}, \eqref{1-p_kl_inequality} and \eqref{log_inequality}, thus guaranteeing the ascent property of ELBO.
\begin{equation}\label{surrogate}
	\begin{split}
        &\text{Q }(\boldsymbol{\theta}^{(t)},\boldsymbol{\gamma}^{(t-1)},\boldsymbol{\pi}^{(t)},\boldsymbol{f}^{(t)}; \boldsymbol{\gamma})\\= & \sum_{i = 1}^{n-1}\sum_{j = i+1}^n\sum_{k=1}^{K}\sum_{l=1}^{K}\Bigg\{\Bigg(\gamma_{ik}^2\frac{\gamma_{jl}^{(t-1)}}{2\gamma_{ik}^{(t-1)}}+\gamma_{jl}^2\frac{\gamma_{ik}^{(t-1)}}{2\gamma_{jl}^{(t-1)}}\Bigg)\Bigg[\mathbbm{1}_{e_{ij}\neq 0}\log(p_{kl}^{(t)})+\\&\mathbbm{1}_{e_{ij}= 0}\log(1-p_{kl}^{(t)})\Bigg]+\mathbbm{1}_{e_{ij}\neq 0}\Bigg[\log(f_{kl}^{(t)}(w_{ij}))-\Big(\int_{\mathcal{X}}f_{kl}^{(t)}(u)du-1\Big)\Bigg]\Bigg\} \\
		&+ \sum_{i=1}^{n}\sum_{k=1}^{K}\gamma_{ik}\Bigg[\log(\pi_{k}^{(t)})-\log(\gamma_{ik}^{(t-1)})-\frac{\gamma_{ik}}{\gamma_{ik}^{(t-1)}}+1\Bigg]
    \end{split}
\end{equation}
To maximize \eqref{surrogate}, we solve $n$ separate quadratic programming problems of $K$ variables $\boldsymbol{\gamma}_i$ under the constraints $\gamma_{ik}\geq 0$, $\forall$ $k \in \{1,\dots,K\}$ together with $\sum_{k=1}^K\gamma_{ik}=1$.


\bigskip
\textbf{M-step}: we first maximize the \eqref{ELBO} with respect to $\boldsymbol{\pi}$ and $\boldsymbol{\theta}$. We have a closed-form update for $\boldsymbol{\pi}$ as $\pi_{k}^{(t)} = \sum_{i=1}^{n}\gamma_{ik}^{(t)}/n$. To update $\boldsymbol{\theta}$, we maximize \eqref{ELBO} using the modified Newton-Raphson method with line search to guarantee the ascent property \citep{dennis1996numerical}.
Now, it remains to update block densities $\boldsymbol{f}$. To this end, we use \eqref{local_cond_log_lik} and the approximation of $\log(f_{kl}(.))$. For simplicity, we use a local polynomial approximation such that $\log(f_{kl}(.))$ can be approximated by a low-degree polynomial $\zeta_{kl}$ in a neighborhood of the fitting point $w$ \citep{loader1996local}:
$
    \log(f_{kl}(u))\approx \zeta_{kl}(u-w)=\sum_{m=0}^p\beta_m^{(kl)}(u-w)^m.
$
 With this approximation, we rewrite \eqref{local_phi_approx_cond_log_lik} as
\begin{equation*}\label{local_approx_cond_log_lik}
    \begin{split}
        &\ell(\boldsymbol{\theta},\boldsymbol{\beta},w;\boldsymbol{Y}\mid\boldsymbol{Z})\\=&\sum_{i=1}^{n-1}\sum_{j=i+1}^n\Bigg\{\big[\mathbbm{1}_{e_{ij}\neq 0}\log(p_{z_i z_j})+\mathbbm{1}_{e_{ij}= 0}\log(1-p_{z_i z_j})\big]+\mathbbm{1}_{e_{ij}\neq 0}\Bigg[K_h(w_{ij}-w)\\
        &\zeta_{z_i z_j}(w_{ij}-w)-\Big(\int_{\mathcal{X}}K_h(u-w)\exp(\zeta_{z_i z_j}(u-w))du-1\Big)\Bigg]\Bigg\},
    \end{split}
\end{equation*}
and then derive the corresponding ELBO as
\begin{equation}\label{ELBO_approx_cond_log_lik}
    \begin{split}
        &\text{ELBO}(w;\boldsymbol{\theta},\boldsymbol{\beta},\boldsymbol{\gamma})\\
		= & \sum_{i = 1}^{n-1}\sum_{j = i+1}^{n}\sum_{k=1}^{K}\sum_{l=1}^{K}\Bigg\{\gamma_{ik}\gamma_{jl}\Bigg(\big[\mathbbm{1}_{e_{ij}\neq 0}\log(p_{kl})+\mathbbm{1}_{e_{ij}= 0}\log(1-p_{kl})\big]+\mathbbm{1}_{e_{ij}\neq 0}\\
        &\times\Big[K_h(w_{ij}-w)\zeta_{kl}(w_{ij}-w)-\Big(\int_{\mathcal{X}}K_h(u-w)\exp (\zeta_{kl}(u-w))du-1\Big)\Big]\Bigg)\Bigg\} \\
		&+ \sum_{i=1}^{N}\sum_{k=1}^{K}\gamma_{ik}\Big[\log(\pi_{k})-\log(\gamma_{ik})\Big]
    \end{split}
\end{equation}
We maximize \eqref{ELBO_approx_cond_log_lik} over a sequence of grid points to estimate the block densities $f_{kl}$.

\textbf{Remark 3.} We note that, the estimated densities in the M-step directly affect the optimization over variational parameters $\gamma$, which can seen by the ELBO constructed after the variational approximation. Given that cluster memberships are usually estimated by the hard clustering over these parameters, weights also affect the clusters.

\section{Simulation Studies}
In this section, we conduct simulation studies to examine our proposed non-parametric methods. The general procedure that we adopt to simulate entails the following steps:
\begin{enumerate}
    \item First we simulate the membership indicators for all nodes from multinomial distribution with parameter vector $\pi$ corresponding to uniform mixture proportions.
    \item We simulate the binary adjacency matrix by simulating dyads in the static network given the cluster membership indicators of nodes. While simulating these dyads we use the network parameters with two settings $\theta_{s_1} = (-1, 1)$ and $\theta_{s_2} = (-0.5, 0.5)$. The first setting corresponds to well separated clusters on the basis of density of edges while second setting considers the more extreme case when the clusters are relatively close.
    \item For each node pair with an edge, we simulate the weight on that edge using true distribution with block parameter that depends on their cluster memberships.
\end{enumerate}
We consider two distributions separately: Normal and Gamma. For space consideration, we include all results about Gamma distributions in the supplementary file.

We compare three different model-based clustering methods in each simulation, which are based on binary ERGM, proposed nonparametric weighted ERGM and ``oracle'' parametric weighted ERGM \citep{desmarais2012statistical} with the correct specification of weight distributions. We consider different node sizes from $100$ to $500$ and $100$ repetitions. Before proceeding, we introduce several average metrics to measure clustering and model parameters estimation performance for different simulation settings over 100 replications. First, to assess the clustering performance, we calculate the log of Rand Index (logRI). The measure $\text{RI}(\boldsymbol{z}, \hat{\boldsymbol{z}})$ calculates the proportion of pairs whose estimated labels correspond to the true labels in terms of being assigned to the same or different groups \citep{rand1971objective}. We calculate logRI as,
\begin{equation*}
	\text{logRI} = \log\Bigg({ \frac{1}{\binom{n}{2}}\sum_{i<j}(I\{z_{i}=z_{j}\}I\{\hat{z}_{i}=\hat{z}_{j}\} + I\{z_{i}\neq z_{j}\}I\{\hat{z}_{i} \neq \hat{z}_{j}\})}\Bigg)
\end{equation*}
Next, to assess the performance of the estimators of network parameters  $\boldsymbol{\theta}$, we consider log of square root of the average squared error (logRASE),
\begin{equation*}
	\text{logRASE}_{\boldsymbol{\theta}} = \dfrac{1}{2}\log\Bigg( \frac{1}{K}\sum_{k=1}^{K}(\hat{\boldsymbol{\theta}}_{k} - \boldsymbol{\theta}_{k})^{2}\Bigg)
\end{equation*}
To assess the performance of the density estimation $\boldsymbol{f}$, we consider the Kolmogorov-Smirnov (KS) statistic,
\begin{equation*}
	\text{KS}_{f} = \sup_{w}\mid\hat{\boldsymbol{f}}(w)-\boldsymbol{f}(w)\mid
\end{equation*}

\begin{figure}[!ht]
    \centering
    \includegraphics[width=0.65\textwidth,height=0.045\textheight]{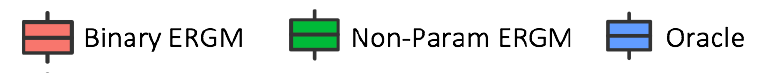}
    \subfigure[$\theta_{s_1}$ with Normal distributions]{
        \includegraphics[width=0.43\textwidth,height=0.28\textheight]{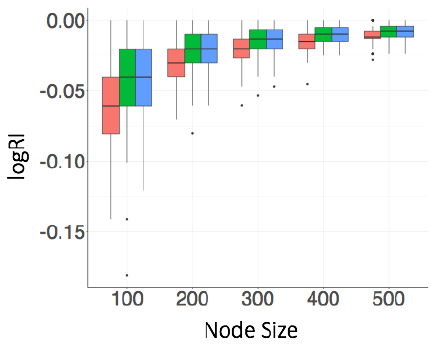}
        \label{fig:RI_s1_normal}
    }
    \subfigure[$\theta_{s_2}$ with Normal distributions]{
        \includegraphics[width=0.43\textwidth,height=0.28\textheight]{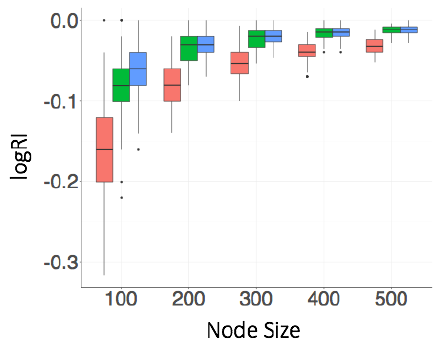}
        \label{fig:RI_s2_normal}
    }
    \caption[Optional caption for list of figures]{Clustering Performance measured using logRI against different node sizes comparing the three models for different sparsity parameter settings under Normal weight distributions}
    \label{fig:RI_normal}
\end{figure}

\begin{figure}[!ht]
    \centering
    \includegraphics[width=0.65\textwidth,height=0.045\textheight]{legend.pdf}
    \subfigure[$\theta_{s_1}$ with Normal distributions]{
        \includegraphics[width=0.43\textwidth,height=0.28\textheight]{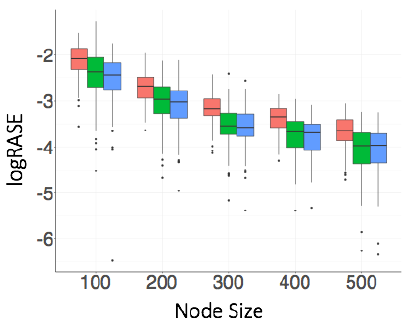}
        \label{fig:RASE_s1_normal}
    }
    \subfigure[$\theta_{s_2}$ with Normal distributions]{
        \includegraphics[width=0.43\textwidth,height=0.28\textheight]{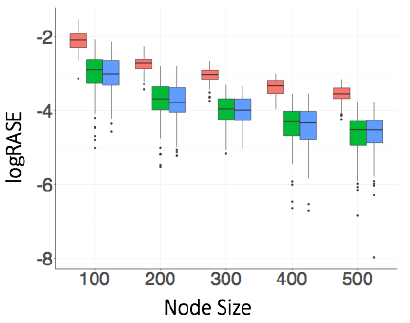}
        \label{fig:RASE_s2_normal}
    }
    \caption[Optional caption for list of figures]{$\theta$ Estimation Performance measured using logRASE against different node sizes comparing the three models for different sparsity parameter settings under Normal weight distributions}
    \label{fig:RASE_normal}
\end{figure}

Based on the metrics defined here, Figures \ref{fig:RI_normal} and \ref{fig:RASE_normal} show clustering and $\theta$ estimation performance for different sparsity parameter settings averaged over 100 simulations of graphs for normal weight distributions. Corresponding figures for Gamma weight distributions look similar and have been moved to Appendix A. The differences in logRI and logRASE for $\theta_{s_1}$ and $\theta_{s_2}$ evidently confirms the expected fact that separating two very close clusters is difficult compared to well separated clusters. It appears that the both the distributions allow a reasonable recovery of the cluster membership indicators, when the graphs considered have more than 100 nodes. As expected, the node size improves the recovery of latent structure and estimation of network parameters $\theta$ in all cases. It can be observed that our proposed nonparametric ERGM outperforms the binary ERGM by a large difference and performs competitively with the oracle method (parametric ERGM with true weight distributions) for all simulation settings. We note that our proposed
strategy is best suited for real world applications when the true distributions for block pairs are unknown.

\begin{figure}[!ht]
    \centering
        \includegraphics[width=0.88\textwidth,height=0.33\textheight]{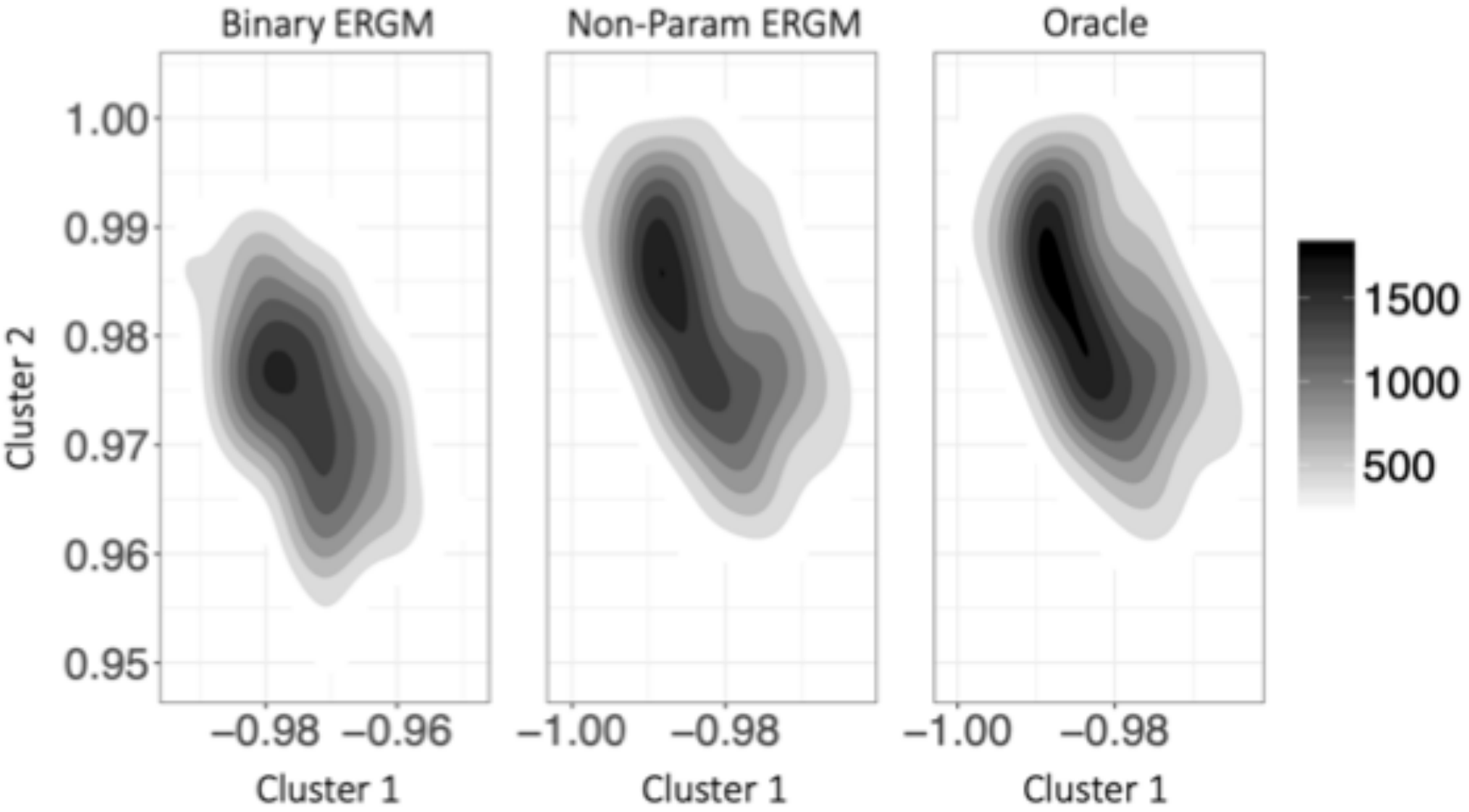}
        \caption{Plots of empirical joint distributions of network parameters $\theta_{s_1}$ for Normal weight distributions over 100 simulations with 500 nodes, comparing the three models for different block distributions}
        \label{fig:theta_contour_s1_normal}
\end{figure}

\begin{figure}[!ht]
    \centering
        \includegraphics[width=0.88\textwidth,height=0.33\textheight]{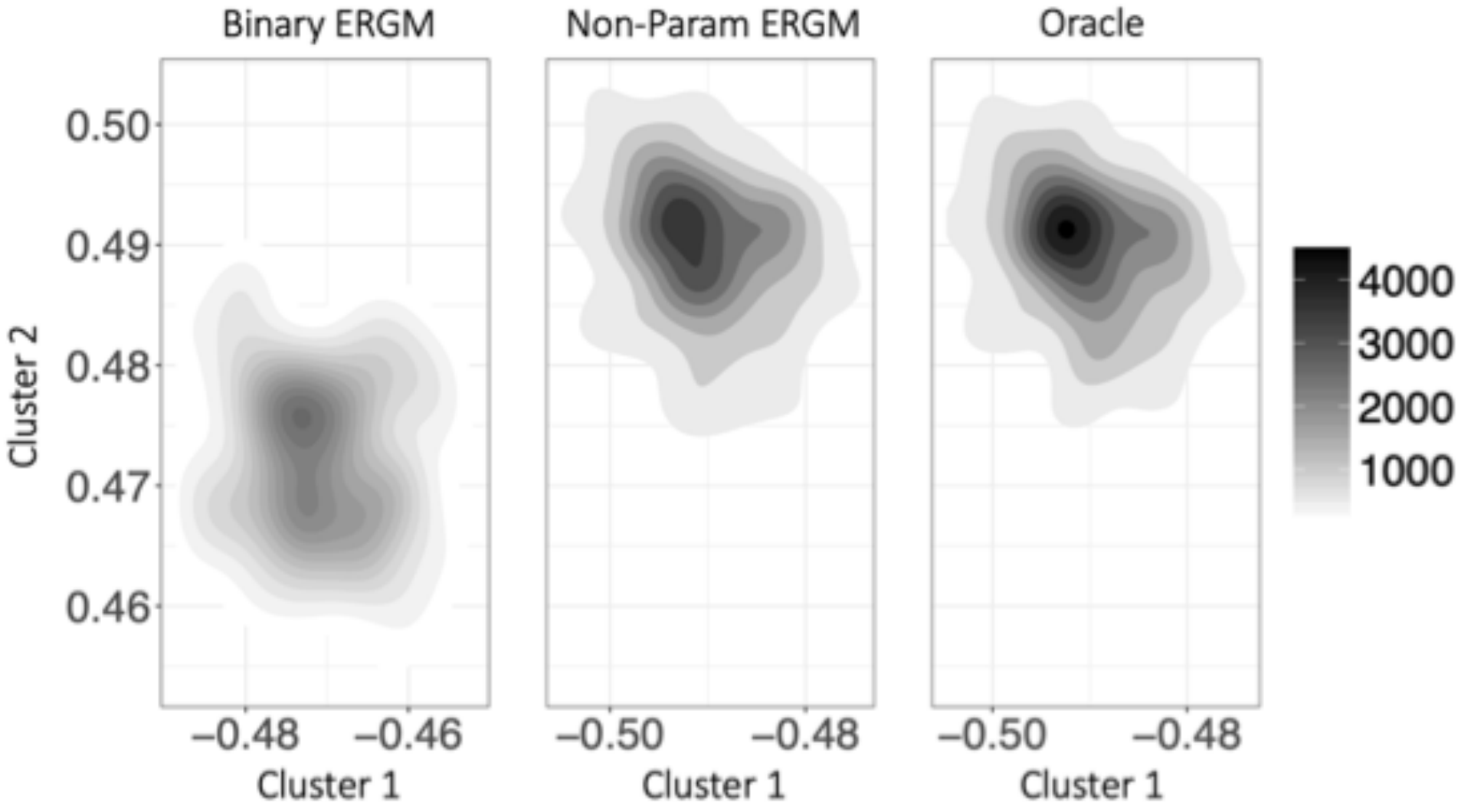}
        \caption{Plots of empirical joint distributions of network parameters $\theta_{s_2}$ for Normal weight distributions over 100 simulations with 500 nodes, comparing the three models for different block distributions}
        \label{fig:theta_contour_s2_normal}
\end{figure}

Figures \ref{fig:theta_contour_s1_normal} and \ref{fig:theta_contour_s2_normal} show the empirical distributions of network parameters $\theta$ for normal weight distributions. Corresponding figures for Gamma weight distributions have been moved to Appendix A. The proposed non-parametric model estimation again outperforms the binary ERGM uniformly for all settings. We also note that the contour plots for the proposed model seem really close to Oracle model, thus demonstrating the power of our approach.

\begin{figure}[!ht]
    \centering
    \includegraphics[width=0.6\textwidth,height=0.025\textheight]{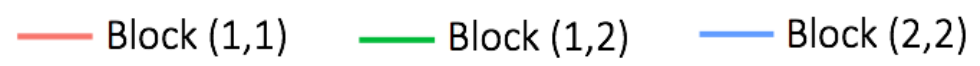}
    \subfigure[$\theta_{s_1}$ with Normal distributions]{
        \includegraphics[width=0.42\textwidth,height=0.26\textheight]{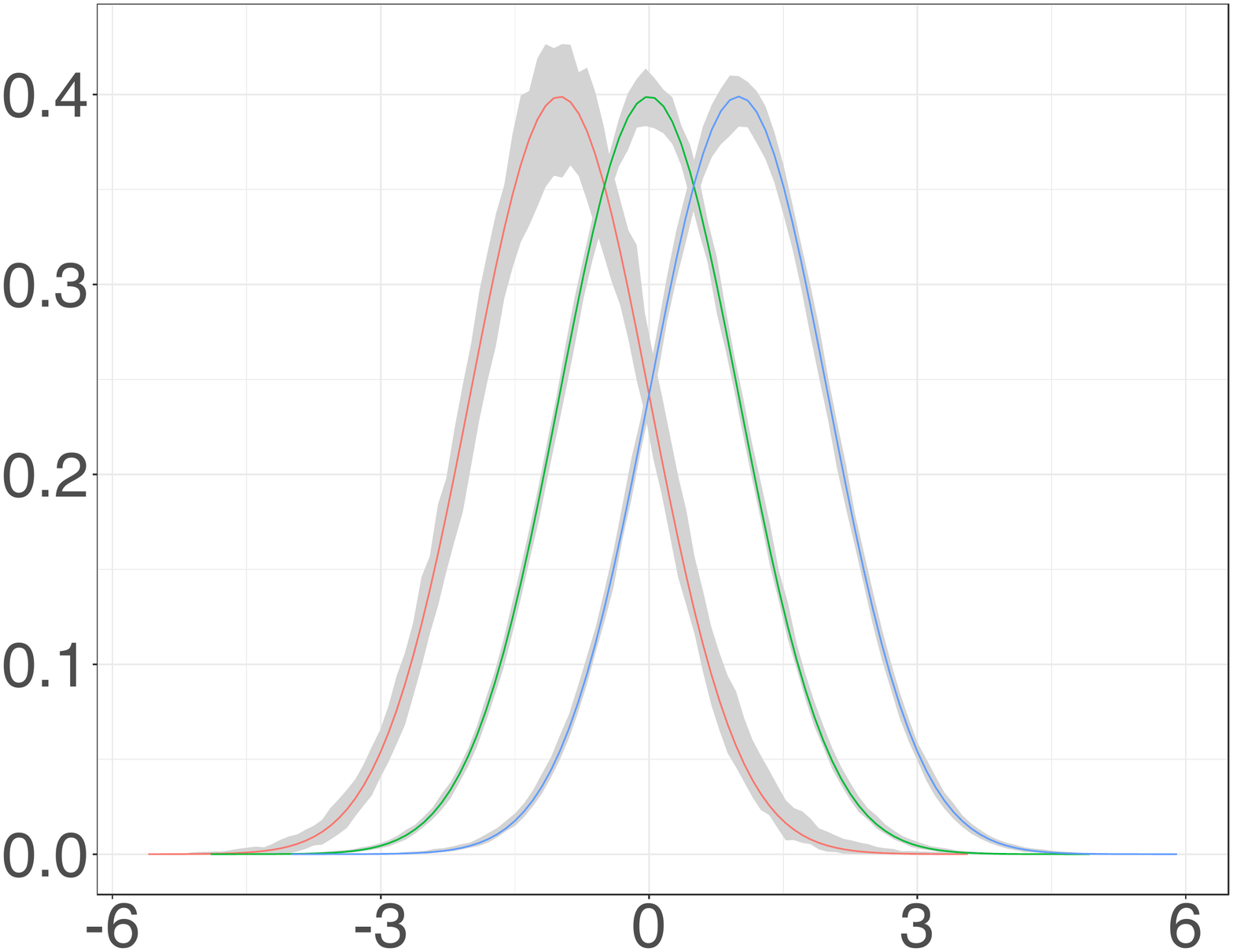}
            }
    \subfigure[$\theta_{s_2}$ with Normal distributions]{
        \includegraphics[width=0.42\textwidth,height=0.26\textheight]{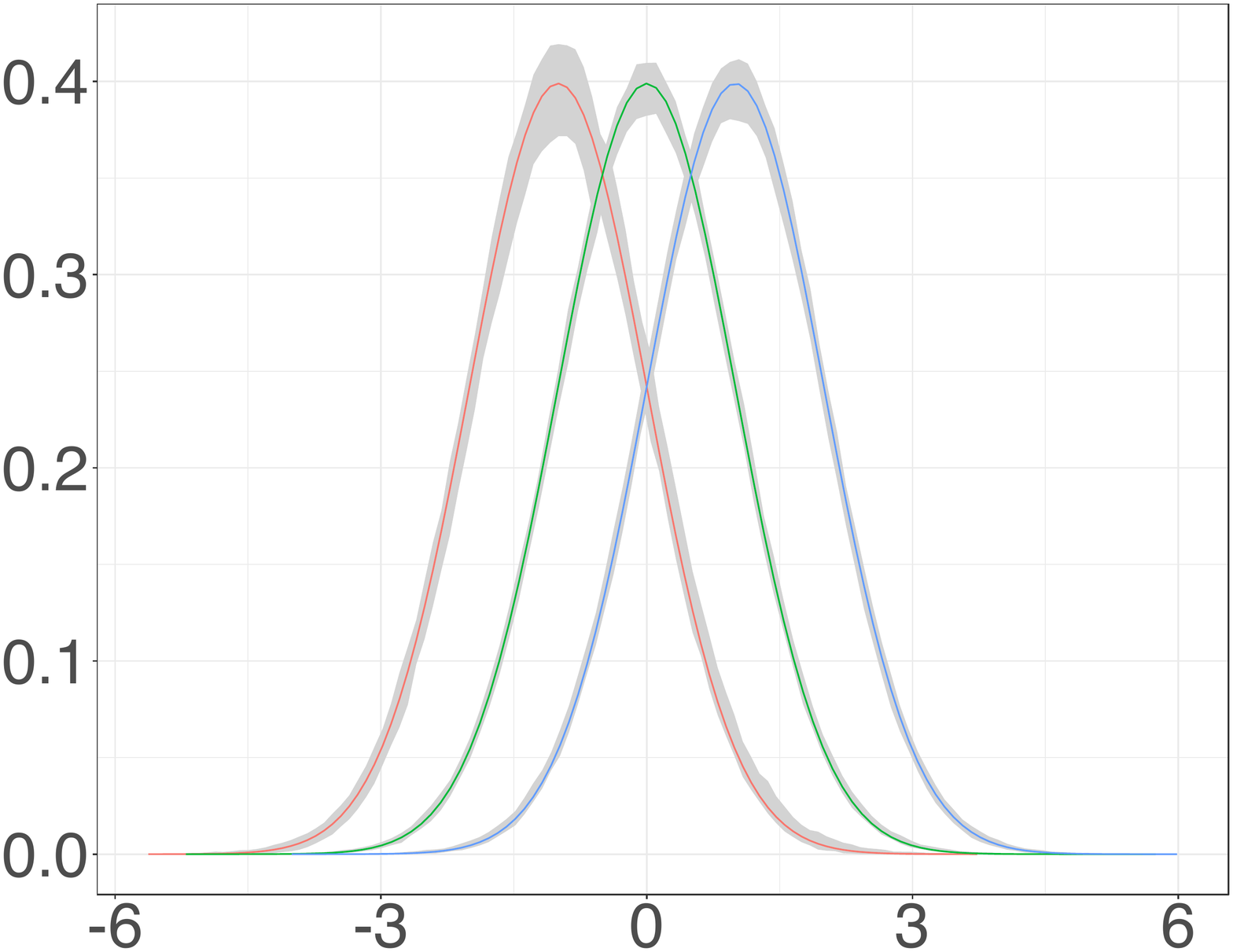}
            }
    \caption[Optional caption for list of figures]{Estimated block densities for normal weight distributions.}
    \label{fig:sim_dens_est_normal}
\end{figure}

Figure \ref{fig:sim_dens_est_normal} shows the estimated block densities within $2.5$ and $97.5$ percentiles for node size of $500$ for normal weight distributions. Corresponding figure for Gamma weight distributions have been moved to Appendix A. Comparing $\theta_{s_1}$ and $\theta_{s_2}$, it is evident that within cluster 1 density estimation improves substantially for $\theta_{s_2}$. This is because cluster 1 is more sparse for $\theta_{s_1}$ compared to $\theta_{s_2}$. Comparing the normal and gamma distributions, we observe that asymmetry of gamma distribution leads to underestimation of within cluster 1 estimated density. Cluster 1 is again most affected since it is most sparse. We point out here that for larger node sizes, these estimated densities will converge to true densities \citep{loader1996local}.

Table \ref{tab:KS} gives the summary of KS statistic for various simulation settings. We note that for sparse cluster 1, comparing $\theta_{s_1}$ and $\theta_{s_2}$, there is a huge improvement when the true distribution is Normal. However the improvement is only minor in case of Gamma due to asymmetry. The differences are much less substantial for other blocks, however Normal uniformly outperforms Gamma for all settings.

\begin{table}[!ht]
\centering
\begin{center}
    \begin{tabular}{lccccc}
\hline
                          &                                                             & \multicolumn{2}{c}{$\theta_{s_1}$} & \multicolumn{2}{c}{$\theta_{s_2}$} \\ \hline
                          & \begin{tabular}[c]{@{}c@{}}Summary\\ Statistic\end{tabular} & Normal           & Gamma           & Normal           & Gamma           \\ \hline
\multirow{2}{*}{Block (1,1)} & Median                                                      & 3.94             & 4.51            & 2.77             & 4.26            \\
                          & Mean                                                        & 4.03             & 4.70            & 2.84             & 4.36            \\ \hline
\multirow{2}{*}{Block (1,2)} & Median                                                      & 1.40             & 1.52            & 1.46             & 1.57            \\
                          & Mean                                                        & 1.41             & 1.56            & 1.49             & 1.61            \\ \hline
\multirow{2}{*}{Block (2,2)} & Median                                                      & 1.51             & 1.53            & 1.63             & 1.76            \\
                          & Mean                                                        & 1.53             & 1.60            & 1.66             & 1.76            \\ \hline
\end{tabular}
\caption{Summary of KS statistic ($\times 10^2$) for the three block densities under different simulation settings for the proposed model, computed over 100 simulations of graphs with 500 nodes}
\label{tab:KS}
\end{center}
\end{table}
\section{Application to Water Pollution Analysis}

In this section, we demonstrate the power of our methodology in an environmental application to study sulfate concentrations in river networks. The dataset consists of three main parts. The first part consists of approximately $865$ sulfate samples measured as concentrations in the units of parts per million (ppm) over several creeks in the Ohio watershed in Pennsylvania. The source include the following online databases: the \href{http://waterdata.usgs.gov/nwis}{USGS National Water Information System}, the \href{http://www.srbc.net/}{Susquehanna River Basin Commission database}, the \href{https://www.
epa.gov/waterdata/storage-and-retrieval-and-water-qualityexchange}{EPA STORET Data Warehouse}, and the \href{www.shalenetwork.org}{Shale Network database} (doi:10.4211/his-data-shalenetwork). The second part consists of the directed geographical river network in the form of locations of all streams and creeks in Pennsylvania. The directions in the network correspond to the actual river flow. The third part consists of $93$ coal mine locations which are suspected to be potential polluters of the river streams posing an environmental risk. Both the latter parts are publically available at \href{http://www.pasda.psu.edu/}{Pennsylvania Spatial Data Access (PASDA)}. We map the latitude and longitude of the sampling sites onto the geographical river network. The 865 sulfate sampling sites become the nodes. Thus the nodes in the network are defined through first part of the data after mapping to nearest streams in the second part. We define the edges according to the path of river flow; for sampling sites A and B, the edge is present when river can flow either from A to B or from B to A directly or through one of its sub-tributary or some intermediary stream. The path information used to define the edges entirely comes from the second part. The spatial weights are constructed as proportional to the strength of influence of the upstream site on the downstream site, measured by difference in concentrations between the sampling locations \citep{peterson2007geostatistical}. For example if the river is flowing from A to B, with measured sulfate concentrations $C_A$ and $C_B$, then weight on the edge between A and B would be defined as $w_{AB}=C_B-C_A$. Note in this definition, the weights could be negative if $C_B<C_A$. In this context, we have transformed the geographically defined 'river network' to a weighted network defined above for the purpose of our analysis.

It is an important question to study the water pollution in river networks. Most of the existing spatial clustering methods rely on some ``neighbourhood" metric that cluster the data points based on their spatial proximity. However, these approaches fail in a river network setup when the two points are very close spatially but still not connected by the river flow or vice-versa. These methods usually rely on some criteria to choose the number of clusters that is heuristic and not rigorously founded on model likelihood. There have been several parametric approaches to study spatial concentration gradients e.g.\ see \cite{lawson2002spatial}. However most methods fail in water pollution applications where the gradients tend to be asymmetric, heavy tailed and multimodal, with an unknown number of modes. We adopt skewness, kurtosis and Hartigan's dip test of unimodality respectively to infer these properties over the whole sulfate network and individual clusters obtained using our model. The results, summarized in Table \ref{tab:dist_properties}, clearly indicate that the gradients follow an asymmetric, leptokurtic and multimodal distribution for all block pairs. This calls for a generalized framework of clustering the weighted network while modeling the concentration gradients without making any distributional assumptions.

\begin{table}[!ht]
\centering
\begin{tabular}{cccc}
\hline
              & Skewness & Kurtosis & \begin{tabular}[c]{@{}c@{}}Hartigan's dip test\\ p value\end{tabular} \\ \hline
Whole network & 0.050    & 3.885    & $< 2.2 \times 10^{-16}$                                               \\
Block (1,1)      & 0.762    & 6.817    & $< 2.2 \times 10^{-16}$                                               \\
Block (1,2)      & -0.011   & 3.447    & $9.37 \times 10^{-6}$                                                 \\
Block (2,2)      & 0.026    & 3.672    & $< 2.2 \times 10^{-16}$                                               \\ \hline
\end{tabular}
\caption{Distributional properties of sulfate concentration gradients}
\label{tab:dist_properties}
\end{table}


The proposed nonparametric weighted network models address all the aforementioned challenges. Now, we apply the proposed method to analyze the sulfate concentration network. In practice, we need to effectively choose the number of clusters (i.e., $K$). Since the likelihood is intractable \citep{biernacki2000assessing}, we follow \citep{daudin2008mixture} to introduce a modified Integrated Classification Likelihood (ICL) criterion:
\begin{equation}\label{icl}
	\mbox{ICL}_{K} = \log P(\boldsymbol{Y}, \hat{\boldsymbol{Z}},\hat{\boldsymbol{f}}) - (K-1)\log n - K\log\left(\frac{n(n-1)}{2}\right),
\end{equation}
where the complete log-likelihood with estimated membership and densities becomes
\begin{equation*}\label{cptlike}
	\begin{split}
		\log P(\boldsymbol{Y}, \hat{\boldsymbol{Z}}, \hat{\boldsymbol{f}}) = & \sum_{i<j}\sum_{k=1}^{K}\sum_{l=1}^{K}\Bigg\{\hat{z}_{ik}\hat{z}_{jl} \Big(\log P_{\boldsymbol{\theta}_{z_{i}z_{j}}}(E_{ij} = e_{ij}\mid\mathbf{Z}=\mathbf{z})\\ & + \log \hat{f}_{z_{i}z_{j}}(W_{ij} = w_{ij}\mid e_{ij}=1,\mathbf{Z}=\mathbf{z})\Big)\Bigg\} + \sum_{i=1}^{n}\sum_{k=1}^{K}\hat{z}_{ik}\log{\hat{\pi}_{k}}
	\end{split}
\end{equation*}
The ICL follows the philosophy of Bayesian model selection criterion. The second term in \eqref{icl} penalizes for the $K-1$ free parameters in the mixture proportions $\boldsymbol{\pi}$. The third term accounts for the penalization of the network parameters \citep{matias2016statistical} given the additive structure of the specified ERGM.

\begin{figure}[!ht]
        \centering
        \includegraphics[width=0.68\textwidth,height=0.52\textheight]{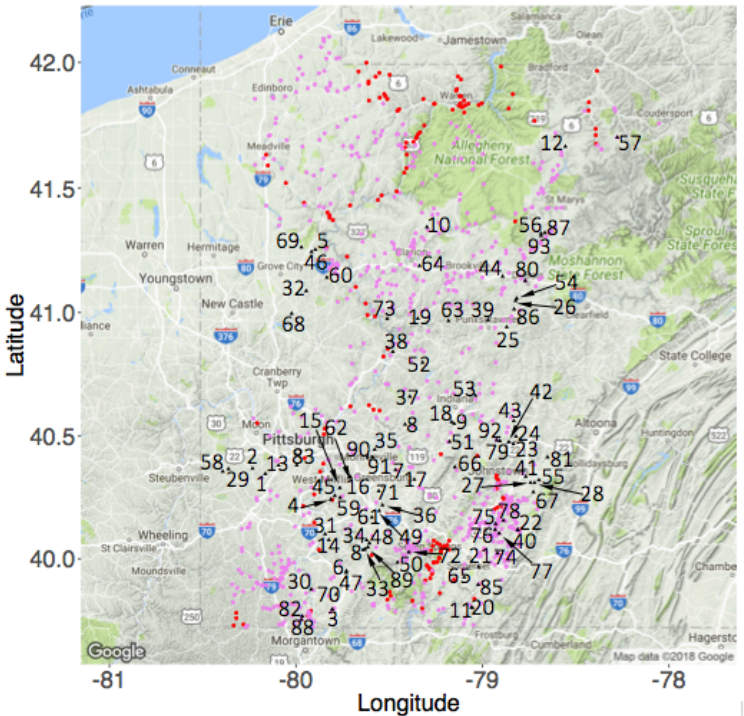}
        \caption{Clustered sulfate sampling sites with $C_1$ in red $\&$ $C_2$ in pink and coal mining sites in black (numbered $1-93$)}
        \label{fig:clustered_river_network}
\end{figure}

We choose the optimal number of clusters by maximizing the modified ICL criterion,
which suggests $K=2$. The estimated network sparsity parameters corresponding to these two clusters labelled $C_1$ and $C_2$ are $-0.521$ and $-2.084$ respectively indicating that $C_1$ has higher degree in terms of the edges $(e_{ij})_{i,j\in C_1}$ on average compared to $C_2$.
We plot the sampling sites belonging to the two clusters overlaying the potential polluter locations in Figure \ref{fig:clustered_river_network}. It can be seen that coal mines $4$, $5$, $15$, $20$, $22$, $38$, $40$, $42$, $45$, $46$ $73$, $75-78$, $82$ and $88$ lie directly either upstream or downstream of nodes in $C_1$ suggesting they may significantly affect the sulfate concentrations. As show in Table \ref{tab:summary_clusters}, $C_1$ consists of relatively more hubs with higher degree on average while $C_2$ consists more nodes with lower degree on average.


\begin{table}[!ht]
\centering
\begin{tabular}{ccc}
\hline
Summary& $C_1$ & $C_2$ \\ \hline
Number of nodes                                              & 147       & 718       \\
Average Degree                                               & 18.72     & 5.21      \\
Minimum                                                      & -385.50   & -389.70   \\
1st quantile                                                & -2.00     & -35.99    \\
Median                                                       & 8.081     & 10.24     \\
Mean                                                         & 40.41     & 23.66     \\
3rd quantile                                                & 66.96     & 111.80    \\
Maximum                                                      & 436.40    & 441.00    \\ \hline
\end{tabular}
\caption{Summary of nodes, degree and edge weights of the two clusters.}
\label{tab:summary_clusters}
\end{table}

\section{Discussion}
We introduce a new nonparametric model-based approach for clustering large-scale weighted networks. The ERGM specification allows the flexibility to incorporate interesting network statistics and the nonparametric density function provides the robustness to study the network weights. We illustrate the power of our proposed method in a real application to study water pollution networks.

In general, our proposed method does not require a parametric specification of network weights and thus it is robust to the model mis-specification, and it can be extended to incorporate the nonparametric mixture functions or parametric constraints \citep{desarbo2017parametric,lee2018nonparametric}. Like most nonparametric methods, when the sample size is limited, our proposed method may not perform well. The semiparametric extension such as \cite{xue2012regularized,xue2014rank} and \cite{fan2016multitask} seems a promising alternative to the proposed nonparametric method. Moreover, the proposed method could be computationally intensive when the number of nodes is huge. To make the proposed methods scalable, we shall follow the stochastic variational methods \citep{hoffman2013stochastic} to employ the minibatch sampling scheme.

\bibliographystyle{apalike}

\end{document}